# An Electronic Measurement of the Boltzmann Constant

Samuel P. Benz[1], Alessio Pollarolo[1,2], Jifeng Qu[3], Horst Rogalla[1,4], Chiharu Urano[1,5], Weston L. Tew[6], Paul D. Dresselhaus[1], and D. Rod White[7]

1 National Institute of Standards and Technology (NIST), Boulder, CO USA;

2 Politecnico di Torino, Italy;

3 National Institute of Metrology (NIM), China;

4 Department of Applied Physics, Univ. Twente, The Netherlands;

5 National Metrology Institute of Japan (NMIJ) Tskuba, Japan;

6 NIST, Gaithersburg MD USA;

7 Measurement Standards Laboratory (MSL), Lower Hutt, New Zealand

## Abstract
The Boltzmann constant was measured by comparing the Johnson noise of a resistor at the triple point of water with a quantum-based voltage reference signal generated with a superconducting Josephson-junction waveform synthesizer. The measured value of $k = 1.380651(18) \times 10^{-23}$ J·K$^{-1}$ is consistent with the current CODATA value and the combined uncertainties. This is our first measurement of $k$ with this electronic technique, and the first noise thermometry measurement to achieve a relative combined uncertainty of 13 parts in $10^6$. We describe the most recent improvements to our Johnson Noise Thermometer that enabled the statistical uncertainty contribution to be reduced to seven parts in $10^6$, as well as the further reduction of spurious systematic errors and EMI effects. The uncertainty budget for this measurement is discussed in detail.

## 1. Introduction
For ten years, NIST has been developing a Johnson noise thermometer (JNT) with a quantized voltage noise source (QVNS) as a voltage reference. The original goal was to improve thermodynamic temperature measurements through quantum-based electrical measurements and to demonstrate the technique through measurements of the triple point of water and the melting point of gallium [1. Following these proof-of-principal measurements, higher temperature measurements were made of the freezing points of zinc and tin, which showed agreement with archival gas and radiation thermometry measurements 2. As a result of these measurements and further research and development [3-10], the QVNS-JNT system evolved through many incremental improvements to the point where a useful measurement of the Boltzmann constant has become practical [11].

The presently accepted value of the Boltzmann constant, 1.3806504(24)×10$^{-23}$ J·K$^{-1}$, is dominated by a single gas-based thermometry measurement with a relative standard uncertainty of 1.8×10$^{-6}$ [12]. There has been much interest in reproducing this measurement, as well as in methods based on different physical principles that might achieve comparable uncertainty [13]. The QVNS-JNT measurement of the Boltzmann constant is quite different from gas-based measurement techniques, in that it is a purely electronic approach that links the SI kelvin to quantum-based electrical measurements. The QVNS, which is a low-voltage realization of the Josephson arbitrary waveform synthesizer, is programmed to produce multi-tone pseudo-noise voltage waveforms with small (<1 μV peak) amplitudes [10]. The Josephson junctions in the QVNS produce voltage pulses with time-integrated areas perfectly quantized in integer values of $h/2e$, where $h$ is the Planck constant and $e$ is the electron charge. The synthesized voltage is intrinsically accurate because it is exactly determined from the known sequence of pulses, the clock frequency, and fundamental physical constants [1, 3]. The QVNS-JNT measurement can, therefore, provide a unique contribution to CODATA analyses of the Boltzmann constant value as well as the planned redefinition of the kelvin [14].

Our QVNS-JNT measurement determines the ratio between the Boltzmann and Planck constants, $k/h$, by matching the electrical power of a synthetic signal and the thermally generated noise power of a resistor at the triple point of water [3, 11]. The Johnson noise of a resistor $R$ at a temperature $T$ defines the thermal noise power, which is characterized by its mean square voltage $\langle V^2 \rangle = 4kTR\Delta f$, where $\Delta f$ is the measurement bandwidth [15, 16]. This relation is an approximation to the Nyquist equation and is accurate to a part in 10$^6$ for frequencies below 10 MHz and temperatures above 250 K. For our measurement of $k$, we measure the voltage noise of a 100 Ω resistor at the triple point of water 273.16 K. The resulting noise voltage has a spectral density of only 1.23 nV/Hz$^{1/2}$, and must be measured with a low-noise cross-correlation technique [17, 18] in order to extract the resistor signal from the uncorrelated amplifier noise, which has a comparable amplitude and spectral density.

The pseudo-noise-voltage density synthesized by the QVNS waveform, $V_Q = (h/2e)QN_J(Mf_s)^{1/2}$, is determined by defined fundamental constants ($h$ and $e$), the number of Josephson junctions $N_J$, the integer number of bits, $M$, that determine the length of the digital waveform, the clock frequency $f_s$, and a constant $Q$. $Q$ is the programmable dimensionless fraction of the peak voltage of the Josephson junction and exactly sets the amplitude of the synthesized waveform.

The Boltzmann constant is determined from values of the Planck constant, the water-triple-point temperature, the measured resistance, the QVNS synthesis parameters, and the measured ratio of the resistor and QVNS noise power ($V^2$) signals [10, 11]:

$$k = \left(\frac{\langle V_R^2 \rangle}{\langle V_Q^2 \rangle}\right)_f \frac{hQ^2 N_J^2 f_s M}{16 T X_R}, \qquad (1)$$

where $X_R$ is the measured resistance expressed in units of the von Klitzing resistance $R_K \equiv h/e^2$. The cross-correlated electrical and thermal noise powers are compared over discrete frequency intervals centered at the frequency of the harmonic tones of the synthesized waveform [10], a process that will be described below in more detail. Since the relative standard uncertainty for the Planck constant is about two orders of magnitude smaller than that for the Boltzmann constant, we use the 2006

CODATA value, $h = 6.62606896(33)\times 10^{-34}$ J·s [12], to determine our value of the Boltzmann constant.

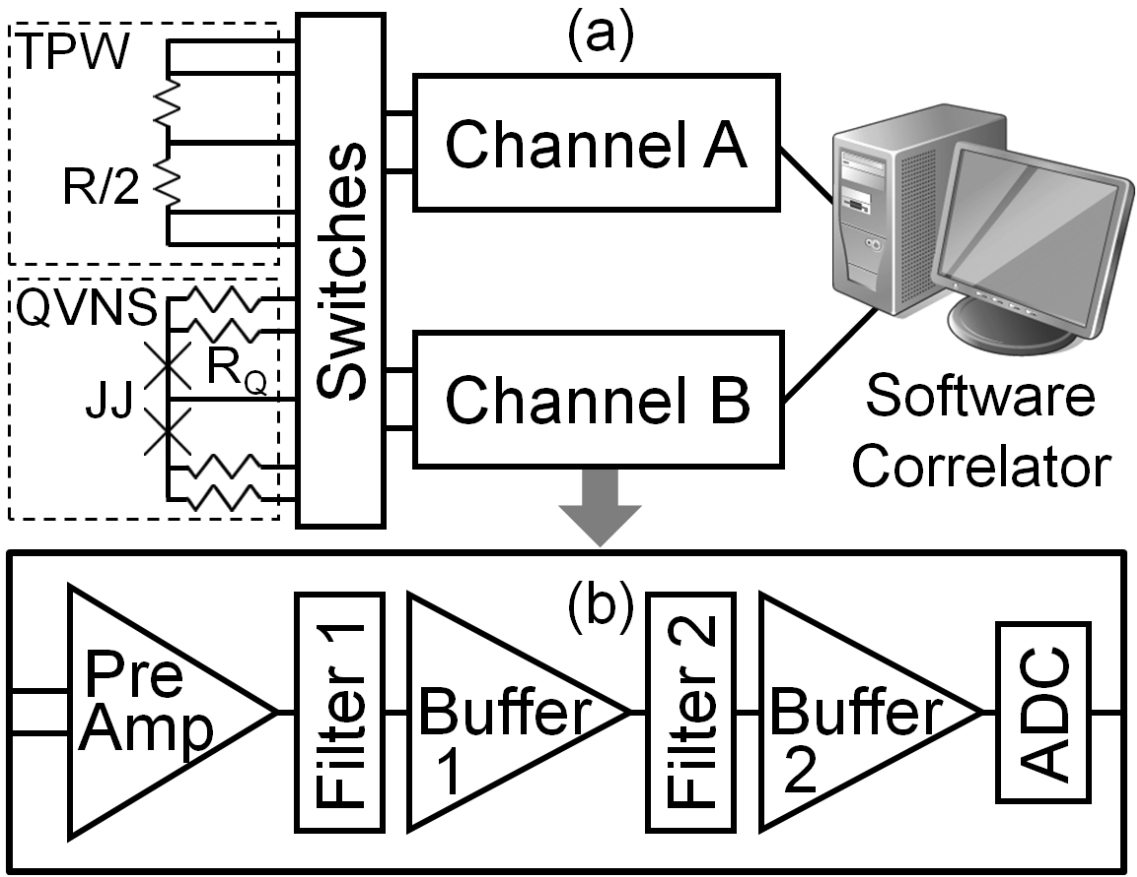

Figure 1. Schematic of the two-channel QVNS-JNT cross-correlation electronics, showing (a) the output wiring of the triple point of water (TPW) and QVNS probes (where the fifth wire is ground), and (b) the specific components of each channel, including the differential input leads to the preamplifier and the optical interface between the analog-to-digital converter (ADC) and the computer that performs the correlation analysis.

## 2. Experimental Apparatus

The circuit schematic in Fig. 1 shows the two channels of the correlator (A and B) that simultaneously measure one of the two voltage sources. The switching network alternates or "chops" between the two input signals at 100 s intervals, such that both channels simultaneously measure either the resistor noise or the QVNS voltage signals. This technique minimizes the effects of time-dependent variations in the response of the cross-correlation electronics. Each channel consists of a series of amplifiers and filters, followed by an analog-to-digital converter (ADC). The digitized signals from each channel are optically transmitted to the computer, where software computes and averages the auto- and cross-correlation spectra of many "chops".

The design of the resistor and QVNS circuits are described elsewhere in detail [1, 2, 10]. The value of the 4 K, on-chip, resistor in each QVNS output lead, $R_Q$, is equal to the total sense resistor value, $R$. The signals from both sources are connected to each preamplifier via a differential twisted pair of leads, and a fifth shield wire that defines ground potential. The resistor circuit is a custom Ni-Cr-alloy foil on an alumina substrate mounted in a hermetically sealed package. The resistor sensor package is mounted on a beryllia header in a custom-made probe designed to reduce thermal errors when inserted into the triple-point-of-water cell. The triple point cell environment is maintained near 0.01 ºC with a thermoelectric cooler.

The QVNS chip is a niobium superconducting integrated circuit fabricated at NIST in Boulder, and has a total of eight Josephson junctions [10]. The chip is mounted on a microwave compatible flexible cryopackage, within a magnetically shielded cryoprobe cooled to 4 K in a 100 l liquid helium storage dewar.

The impedance of the transmission lines, which transmit the voltage signals between the preamplifier in a given channel and each source, are carefully matched in order to maximize the measurement bandwidth and minimize frequency-dependent correction terms in the analysis [3]. Ferrite beads are used on the differential leads at the amplifier input to suppress Colpitts' oscillations that can arise in the JFET cascode amplifiers at frequencies above 10 MHz.

Each high-performance JFET preamplifier has low input noise voltage (~0.85 nV/Hz$^{1/2}$), high gain (~3160× or 70 dB), high linearity, and high common-mode rejection (~100 dB at 100 kHz), over the full measurement bandwidth. The preamplifier also operates without feedback around the JFET input stage in order to prevent errors from correlations between the input noise voltage and the input noise current [19]. In addition, the preamplifier also provides a high input resistance to minimize the load on the resistor noise source. A significant portion of the research and development of the QVNS-JNT system has entailed optimizing the preamplifier and improving these features, especially the linearity [5-9].

Two passive low-pass filters, and corresponding buffer amplifiers, each with 11× gain, are used in each channel [8-9]. Both filters have 11 poles, but have different cut-off frequencies of 650 kHz and 800 kHz. This large effective number of poles is necessary to maximize the measurement bandwidth and ensure that aliased signals from the ADCs are at least 120 dB lower than the "in-band" signals for frequencies up to 650 kHz. This steep amplitude reduction prevents the aliased high-frequency signals of each channel from adding to the measurement of the in-band signals.

The ADCs sample the voltage signals for nearly 1 s with $2^{21}$ samples at a rate of 2.083 MSa/sec (50 MHz clock for 24 bits, of which only 16 are used for data [9]). This produces frequency resolved bins of 0.9934 Hz and a 1.042 MHz Nyquist frequency that defines the ADC bandwidth. The 1 s data sequences from both channels are optically transmitted to the digital receiver card in the computer, which computes the fast-Fourier transform (FFT) of the sequences and calculates the auto-correlation for each channel and the cross-correlation between the channels. The receiver accumulates and averages 100 of the 1 s FFTs for each "chop", and stores the two real and one complex frequency-domain data arrays on the computer. The switching network then "chops" the correlator input to the alternate source for the next measurement.

The QVNS synthesized pseudo-noise voltage spectral density is set to exactly $V_Q \equiv 1.228000$ nV/Hz$^{1/2}$, which closely matches the noise voltage spectral density $V_R = 1.228267$ nV/Hz$^{1/2}$ of the 100.0051 Ω resistor at the triple point of water. The QVNS pseudo-noise waveforms consist of a "comb" of harmonic tones that are equally spaced in frequency (typically odd consecutive tones) and have identical amplitudes and random relative phases [4, 10]. The spacing of the harmonic tones $\Delta f_h$ is determined by the code length and the choice of harmonics.

Data presented in this paper are from two different QVNS waveforms that both consist of odd harmonic tones, but have different code lengths, $M = 3\times2^{23}$ (~25 Mb) and $3\times2^{26}$ (~200 Mb), and correspondingly different tone spacing ($\Delta f_h = 2f_s/M$) of 794.73 Hz (25 Mb) and 99.34 Hz (200 Mb). Since the codes' lengths differ by a factor of eight, the rms voltage amplitude of the synthesized tones is also different for each waveform, 34.61846 nV (25 Mb) and 12.23948 nV (200 Mb). Although the code lengths, tone densities, and tone amplitudes are different for the two waveforms (requiring them to have different values of $Q$), the total voltage noise $V_Q$, and the average voltage spectral density of each waveform remain the same.

## 3. Experimental Results

Below we present the results from the analyses of two different data sets. The first, data set 1, used the shorter 25 Mb-length QVNS waveform, while the second, data set 2, used the 200 Mb waveform. Measurements of both of these data sets benefited from several recent improvements to the experiment. One of the improvements was the use of mu-metal shielding around the electronics to reduce electromagnetic interference (EMI) [9]. Some EMI signals were found to cause the preamplifiers to occasionally overload, which necessitated the removal of some chop sequences from the analysis.

Newer improvements, which also helped reduce "EMI-induced amplifier overloads," include better management of ground connections with a low-resistance connection to a newly installed earth point, elimination of ground loops, and dedicated mains power lines and conduits. Figure 2 shows how the intermittent larger autocorrelation amplitudes of individual chop measurements corrupted by overload signals were found in data set 1 by characterizing an individual frequency bin over many chops [9]. The improvements sufficiently reduced the EMI, such that the resistor measurements showed no overloads (compared to [9]) and QVNS measurements only showed occasional overloads in channel B. These remaining overloaded chops were removed in the software analysis by eliminating those eight "chops" and the corresponding resistor chops from the accumulated data.

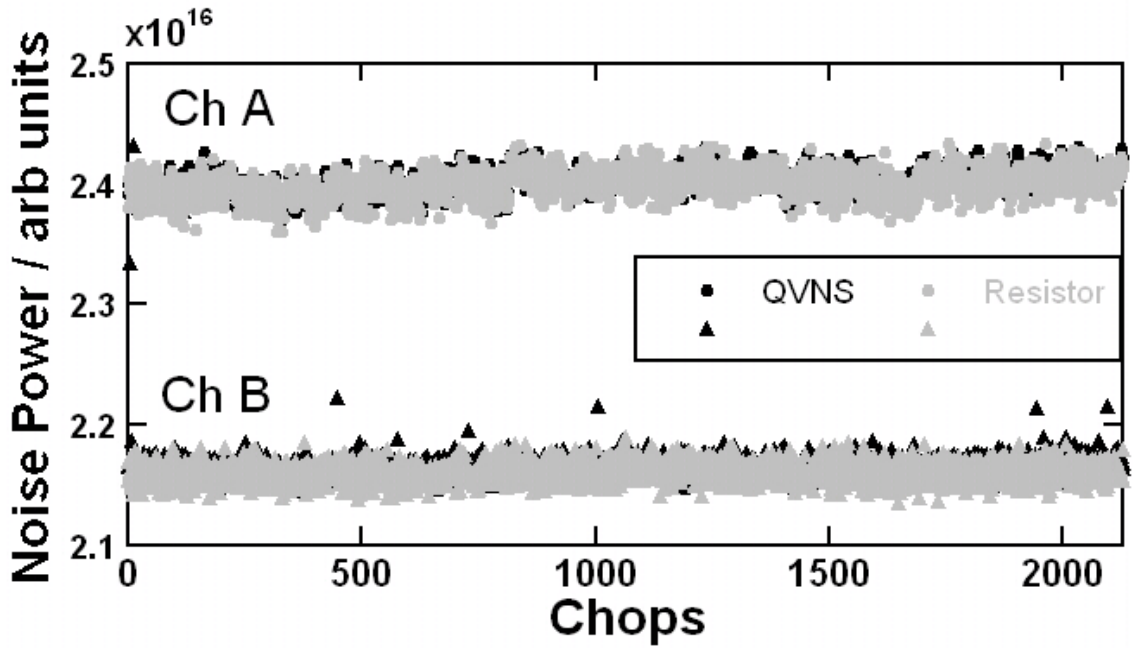

Figure 2. Auto-correlated resistor and QVNS signals of both channels for the 1.589 kHz bin (1 Hz wide) from data set 1 vs. the chop number (consecutive 100 s intervals). Eight chops of the Channel B QVNS measurements indicate an overload event.

Increased linearity of the preamplifiers was another improvement. This was achieved through the addition of an active circuit that nulls the dc offset at the input of the differential stage of each preamplifier. In previous measurements, the dc offsets for the different preamplifiers were found to be directly proportional to the measured amplifier nonlinearity [8]. This is consistent with the imperfect matching of operating points of the transistors in the differential input stage, which leads to an offset accompanied by incomplete cancellation of the even-order distortion effects. As a result of the improved preamplifier linearity, the dominant nonlinearity in the measurements now appears to occur in the ADC circuit and its on-board amplifier [8].

For both data sets, the cross-correlation electronics is powered with Li-ion batteries. However, we have not implemented continuous recharging because the measured statistical uncertainty increases intermittently for some chop measurements. This appears to be similar to the EMI-induced amplifier overloads, but it is a more subtle and difficult effect to detect and eliminate. Therefore, the data presented in this paper has the electronics on isolated Li-ion batteries. The batteries are recharged during the day and the data is collected overnight.

The averaged cross spectra for the signals from both the QVNS and resistor sources for both data sets are shown in Fig. 3. The data show the frequency response of the measurement system, and in particular the combined transfer function of the filters, which dramatically decrease the signals for frequencies above 650 kHz (by seven decades at 1 MHz. There are a number of important features that can be seen from these data. The FFTs of the resistor signals for the two data sets are similar, but the amplitudes of the QVNS harmonic tones differ by eight-fold for the two data sets and differ from the resistor power by their respective tone spacing $\Delta f_h$. As a result of averaging the cross-correlated signals over many samples, the *uncorrelated* noise floors of both QVNS spectra are more than two-orders of
6

magnitude below those of the *correlated* resistor noise powers, even though the resistor and preamplifier noise powers are of comparable value. Note that the QVNS harmonic tones at frequencies above the Nyquist frequency are aliased back to lower frequencies not coincident with the lower frequency tones, while the aliased resistor noise signals are combined with the non-aliased signals in root-sum-square fashion to produce larger rms amplitudes.

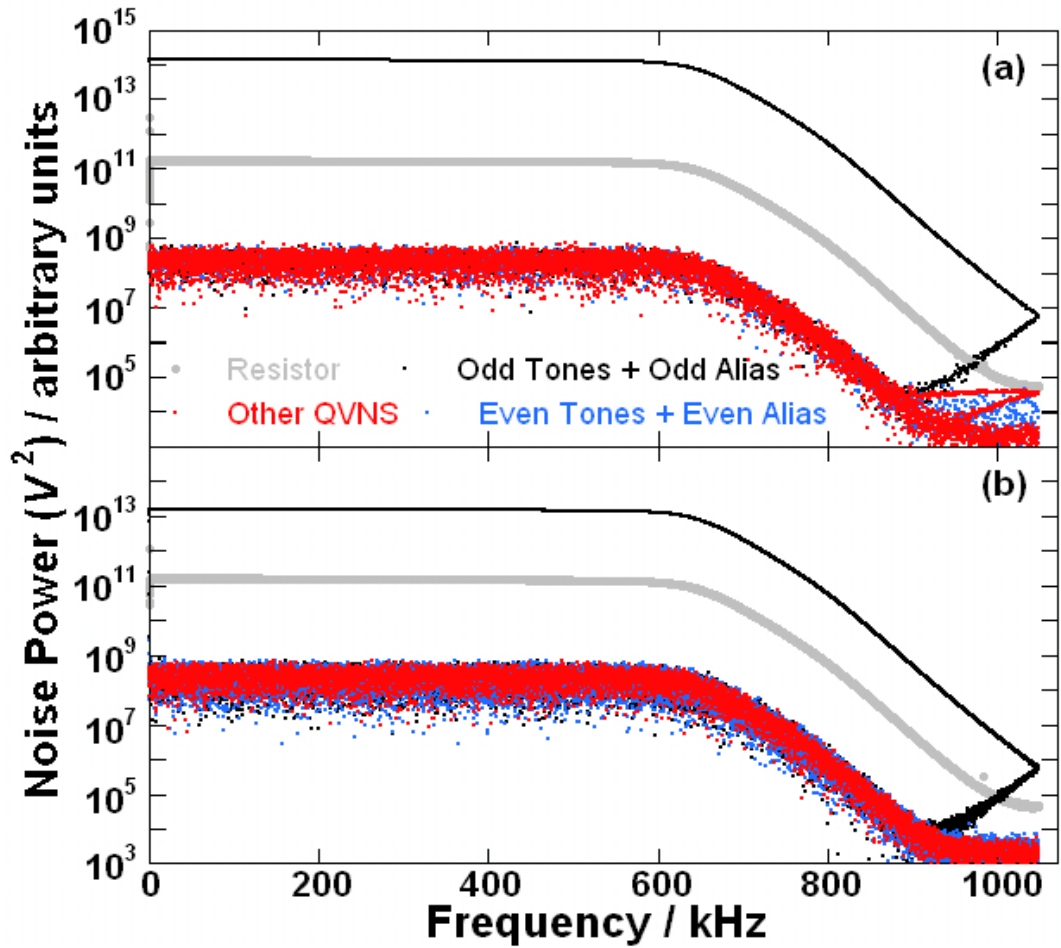

Figure 3. Cross-correlated measured noise power signals for data sets (a) 1 (794 Hz tone spacing) and (b) 2 (99 Hz spacing), showing the QVNS-synthesized comb waveform (black and red) and the resistor noise (gray). The black points represent the QVNS comb tones while the red points show the noise floor after averaging for 116.6 h (2098 chops each source) for both data sets.

One of the more distinct differences in the spectra is the absence in Fig. 3(b) of second-harmonic distortion of the QVNS tones (blue data points including their aliases), which is noticeable in Fig. 3(a) at frequencies above 850 kHz. Experimental investigations suggest that this distortion is due to nonlinearities in the ADC circuit (as mentioned above and in [8]). It is not visible in the data of Fig. 3(b), because the same total second-harmonic-distortion power is distributed as 64 times the number of distortion products of 1/64 the amplitude, with random phases, across 8 times as many FFT bins, with amplitudes an average of 27 db (a factor of $8\sqrt{8}$) below those of Fig.3(a).

Another feature that is different for the two data sets is that spurious signals and their aliases can be seen above 880 kHz in Fig. 3(a). The source of these signals was found to be an oscillation in the ADC voltage regulators, which seems to have mixed with the QVNS tones. This oscillation was fixed and does not appear in the spectra of data set 2.

Two additional changes produced minor differences in the frequency responses of both channels for the two data sets. First, the frequency response of the transmission lines for data set 2 had more attenuation in the measurement band than for data set 1, because ferrite beads of different material were used at the four preamplifier inputs. Secondly, while adding the automatic output voltage compensation circuit to the last stage of the differential amplifiers of the preamplifier circuit, a resistor was accidentally left open-circuit and slightly changed the amplification of the non-inverting part of the differential amplifier. This happened in the preamplifiers of both channels prior to the measurements of data set 1. As a result, the preamplifiers had a reduced common mode rejection so the JNT system was more sensitive to EMI. This problem was discovered and repaired prior to the measurements of data set 2. Note that all chops of data set 2 were entirely free of EMI-induced overloads of the preamplifiers and this appears to be due to the elimination of ground loops.

Some of the features described above are also shown in Fig. 4, which plots the power spectra (calculated as the autocorrelation), rather than the cross spectra, of the measured noise power for each channel for both data sets. In order to directly compare them on a similar vertical axis, the noise power of both data sets are summed (or "rebinned") over the same 794 Hz frequency intervals, which is the frequency spacing of data set 1. The difference between the noise powers of each data set is shown in Fig. 4(b) for both channels. The higher sensitivity to external noise for data set 1 is visible in both the noise power and difference data, which show larger power from EMI signals at a few frequencies. The low-frequency noise power for data set 2 is about 9 % lower than that of data set 1, largely the result of the gain shift from the open-circuit resistors. The frequency-dependent difference between the noise powers is clearly not quadratic, which suggests that the response is a combination of impedance matching differences due to the lower-impedance ferrites for data set 1 and sensitivity to the common mode noise signals from both sensors. Unfortunately, we did not characterize the common mode rejection ratio (CMRR) of the open-circuit resistor circuit for data set 1, so this supposition was not confirmed.

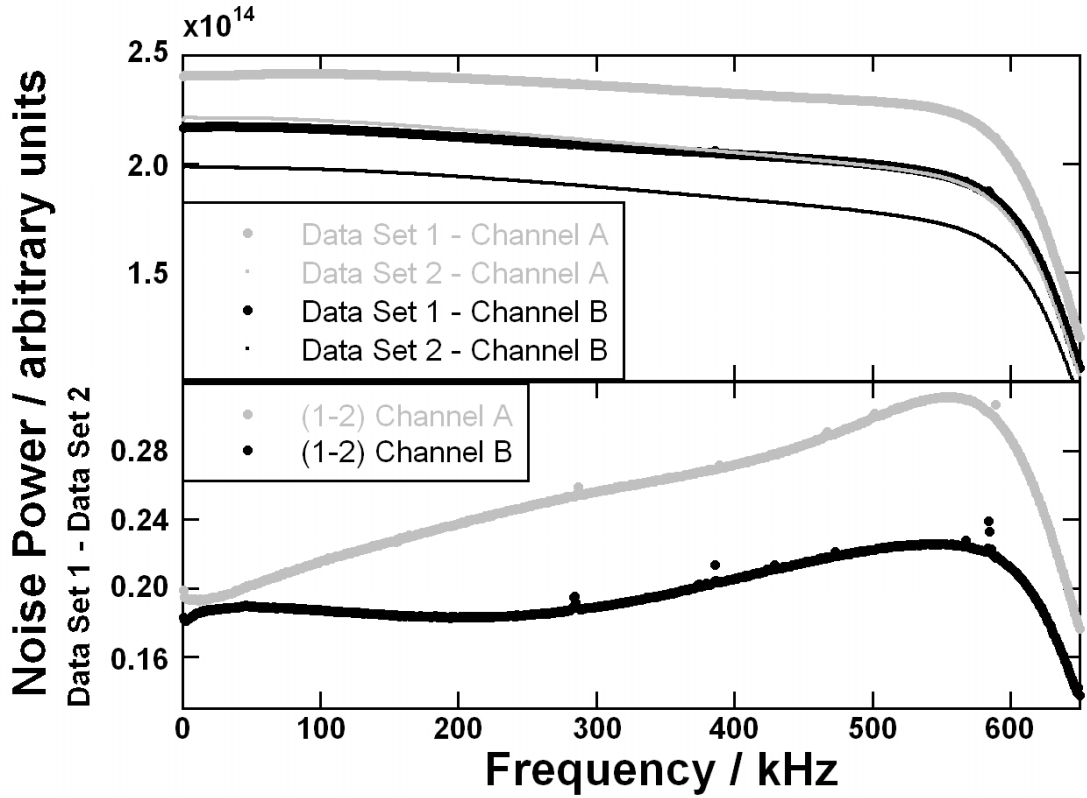

Figure 4. Frequency response of the measured noise power signals for data sets 1 and 2 showing (a) the auto-correlation of each channel and (b) the difference between each channel.

After measuring a large number (of chops) of cross-correlated power spectra for each data set, we average all the data and calculate the noise power over frequency intervals of the harmonic tone spacing $\Delta f_h$ (the effective resolution bandwidth), where the center frequency of these "rebinned" data are at the frequency of each odd harmonic tone [3]. The ratio of the real parts of both the thermal and QVNS noise power is calculated for each of these frequencies. Then, a two-parameter, least-square fit, $a_0 + a_2 f^2$, is used to analyze these ratios over the 650 kHz measurement bandwidth. Fitting the data in this way removes any remaining frequency-dependent differences between the electronic and thermal noise power [9], which result from small differences in the time constants due to imperfect impedance matching. The resulting fitting coefficients, $a_0$ and $a_2$, and the relative standard uncertainty for each coefficient are important characteristics of the data.

Figure 5 shows the residuals of the fitted ratios for both data sets. Because of its smaller frequency range $\Delta f_h$, data set 2 has eight-times more points than data set 1 and the residuals have a correspondingly larger variance. The plot shows that the amplitudes of the residuals of both data sets have negligible frequency dependence over the measurement bandwidth up to the 650 kHz cutoff frequency of the first filter. In fact, these are the "flattest" residuals we have measured with respect to all previous measurements. This flatter frequency response of the electronics is a result of many improvements, but in particular improved amplifier linearity and the reduced aliased signals due to the larger number of filter poles.

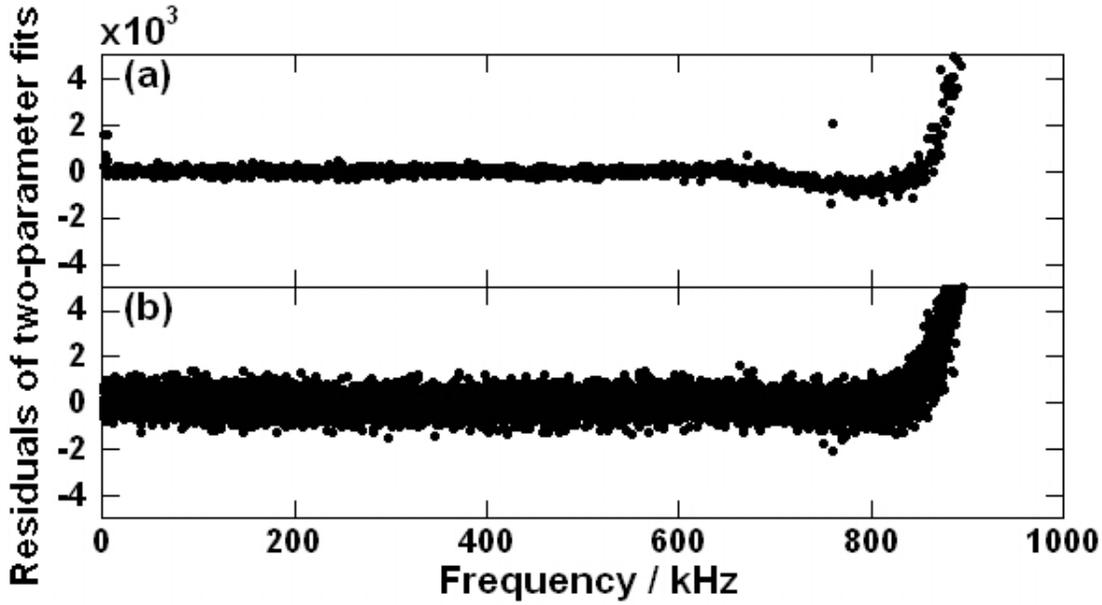

Figure 5. Frequency response of the ratio of the thermal and QVNS cross-correlated noise powers after least square fitting over the 9.934 kHz to 643 kHz frequency range: residuals of (a) data set 1 and (b) data set 2. Each data set was averaged for over 117 hours and includes 2098 chops.

For the Boltzmann constant measurement, we are primarily interested in the coefficients produced from fitting the data over a measurement bandwidth from 10 kHz to 650 kHz ($\Delta f_m$ = 640 kHz). The maximum frequency was chosen to match the cutoff frequency of filter 1. The low frequency starting bin was chosen to avoid frequencies below 10 kHz, which had been susceptible to overload effects. When overloads are not present, then the low frequency starting point is typically reduced to 2 kHz, which is above the 1 kHz high-pass cut-off frequency of the preamplifiers and avoids the difficult-to-shield low-frequency harmonics of the mains supply.

In order to explore the self-consistency of the data, especially the quadratic response, we also analyzed the data over successively smaller bandwidths, with different maximum frequencies. Figure 6(a) shows the difference of the constant "offset" coefficient $a_0$ from the calculated value $a_0^{2006}$= $4kTR/(1.22800002 \times 10^{-9})^2$ = 1.00042923, which is based on the 2006 CODATA value of $k$. The difference expression $a_0 - a_0^{2006}$ is equivalent to the relative difference $(k - k_{2006})/k_{2006}$, which is used below, as well as the relative temperature differences $(T - T_{90})/T_{90}$, which we presented in previous publications. The statistical uncertainty at each bandwidth is based on the standard error estimates for $a_0$ from that fit.

Figures 6(b) and 6(c) show the bandwidth dependence of the second order coefficients for the two data sets. Compared to $a_0$, the $a_2$ coefficients are one order of magnitude smaller for data set 1 and two orders of magnitude smaller for data set 2. These small second-order coefficients demonstrate that the resistor and QVNS transmission lines are well matched, especially for the data set 2 measurements, and that there is no significant quadratic component to the ratio measurement.

For all of these ratio measurements, a quadratic fit was used to determine the noise power ratio and remove any remaining frequency-dependent differences between the electronic and thermal noise power [2,3], which result from small differences in the time constants due to imperfect impedance matching. The offset coefficients for the four largest bandwidth calculations are in excellent agreement for the two data sets, which would suggest confidence for the full bandwidth calculations. However, for the smaller bandwidth calculations, the offsets for the two data sets are in greater disagreement and the offset for data set 2 is much larger than for the larger bandwidth calculations. Another interesting feature is that the $a_2$ coefficients for each data set appear to have an inverse frequency response with respect to that of $a_0$. These unusual features may be real effects of the measurement, such as possible common mode signals for example, or they may be artifacts of our fitting procedure. Additional measurements and analysis will be required to understand these features. We include these effects as a type B uncertainty, in the following uncertainty analysis.

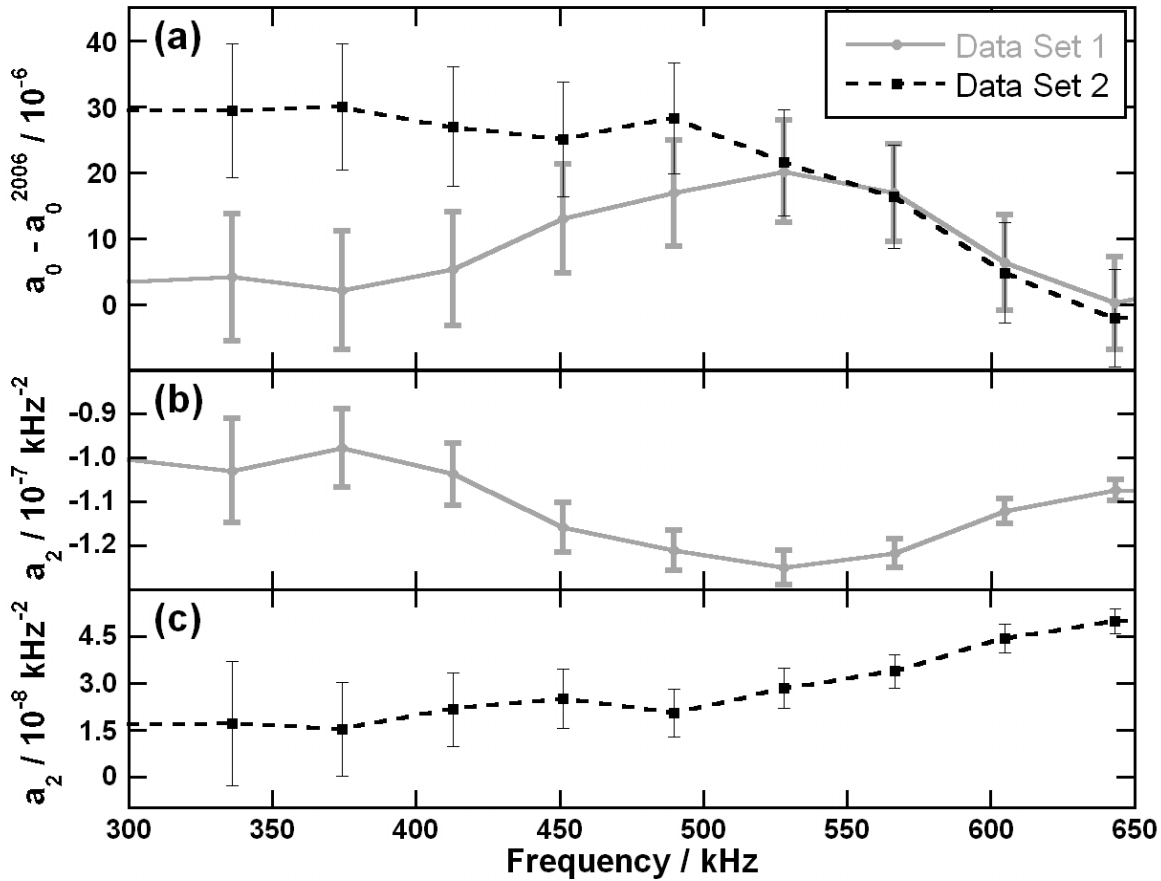

Figure 6. Quadratic fit coefficients determined by fitting the data over different maximum frequencies for both data sets: (a) $a_0$ difference from 2006 constants and $a_2$ for data sets (b) 1 and (c) 2. Connecting lines show the trends.

Finally, in Fig. 7 we present change in the constant offset coefficient $a_0$ for each data set, including the value found from averaging the data from *all* chops, as well as values determined from smaller data sets that were measured on different days. The offset coefficients were determined by fitting the data up to the 643 kHz maximum frequency. An interesting point, which is not shown in Fig. 7, is that the coefficients of data set 2 were not substantially changed when the low frequency starting bin was reduced from 9.984 kHz to 1.986 kHz. However, the coefficients of data set 1 changed significantly, indicating that this data set contained additional chops (that were not manually removed like the largest ones) that are possibly corrupted by EMI.

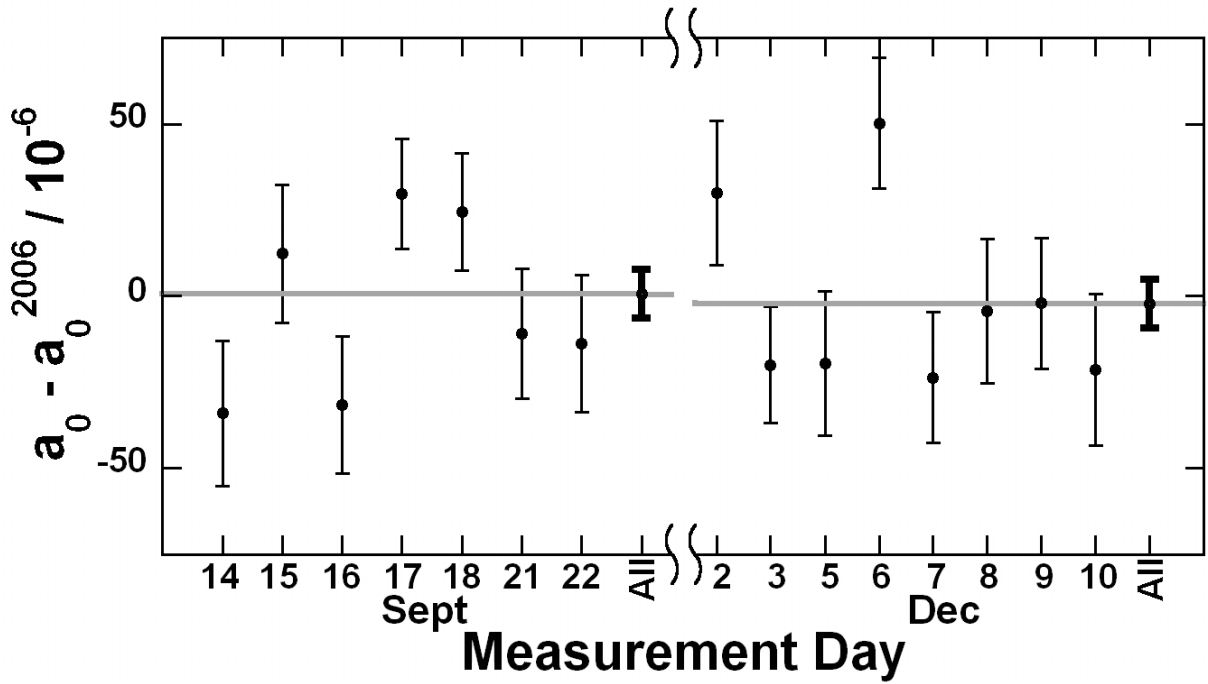

Figure 7. Difference of the constant "offset" coefficient $a_0$ from the calculated value $a_0^{2006}$ determined from both complete data sets (*All*), and for smaller data sets (with fewer chops and shorter times ranging from 14 h to 24 h each) collected on different days. Error bars represent the k=1 relative uncertainty and the fits were determined over the 9.938 kHz to 643 kHz frequency range.

Table I. Difference in the constant "offset" coefficient $a_0$ from the calculated value $a_0^{2006}$ for each data set (fit over 10 kHz to 650 kHz) and for the combined data, including the standard uncertainty.

| Data Set | $(a_0-a_0^{2006})*10^{-6}$ | $\sigma(k=1)$ |
|---|---|---|
| 1 | 2.25 | 7.40 |
| 2 | -1.09 | 7.36 |
| Combined | 0.58 | 7.38 |

## 4. Uncertainty Analysis

The uncertainties are divided into four main categories that correspond to specific factors in Eqn. 1. These are: the measured spectral power ratio (SPR), $\langle V_R^2 \rangle / \langle V_Q^2 \rangle$ ; the measured resistance $R=X_R R_K$; the realization of the water triple point temperature $T_{WTP}$; and the QVNS reference waveform scaling factor $Q$. The tabulated uncertainties are stated as relative standard (k=1) uncertainties, $u_r(x) \equiv u(x)/x$. The uncertainty $u_r(SPR)$ associated with the measurement of the spectral power ratio is by far the dominant uncertainty. Other factors in Eqn. 1 such as those associated with the sampling frequency and the Plank constant have negligible uncertainty.

## 4.1. Spectral Power Ratio

*Noise statistics*

The spectral power ratio is determined for each data set in the low frequency limit via the fit parameter $a_0$. A simple quadratic frequency correction accounts for the degree to which the impedances of the R probe and QVNS probe remain slightly unmatched. The Least-Squares two-parameter fit implements the correction model to determine both $a_0$ and the small quadratic coefficient $a_2$. The statistical uncertainty $u_s(a_0)$ is essentially a measure of the magnitude of the random variations remaining in the cross spectra for the available averaging time $t_a$ [4]. As a statistic, $u_s(a_0)$ is proportional to the standard deviation in the mean for the entire measurement bandwidth of $\Delta f_m = 640$ kHz and is proportional to $(\Delta f_m\, t_a)^{-1/2}$. Values of $u_s(a_0)$ for the two data sets are given in Table I.

*Dielectric loss*

The correction model assumes a pure quadratic frequency dependence in the ratio spectrum and this has a strong physical basis. The leading term in all approximations of the ac characteristics of the both the transmission lines and the preamplifier noise will be quadratic [19]. When approaching uncertainties on the order of $1\times10^{-6}$ however, other forms of frequency dependence need to be considered. The most important of these is a linear frequency term $a_1$ which would be expected due to dielectric loss in any insulators forming shunt admittances of the input circuits.[20] There is no practical way, however, to evaluate such small linear terms in the spectra via simple Least-Squares statistics alone due to the magnitude of the remaining correlated noise. In practice, only the quadratic term can be evaluated via statistical fitting for the noise data presented here.

Instead, we estimate the value of $a_1$ from knowledge of the typical admittances associated with the preamplifier input circuits [2]. The most likely origin of dielectric losses are those originating in the printed circuit board traces for the relays of our switch card made from FR4 composite insulator material [21]. There are three linear frequency terms that arise as each of the two probes couple differently to the input circuit capacitance producing both filtering terms and a correlated noise term. The two filtering terms associated with the dielectric loss are nearly canceled due to the balancing of series termination resistors $R_Q$ in the QVNS probe with the differential resistance $R$ of the resistor probe according to $R_Q \cong R$ (see Fig. 1). This leaves one remaining linear frequency term that cannot be balanced, since it exists only when coupling to the resistor probe and not to the QVNS probe. This leads to a correlated dielectric-noise term from the FR4 given by,

$$a_1 \cong 2\pi \frac{T_{FR4}}{T_{WTP}} R C_{FR4} \tan \delta_{FR4}, \qquad (2)$$

where $C_{FR4}$, $\tan\delta_{FR4}$ and $T_{FR4}$ are the shunt capacitance, loss tangent, and temperature of the circuit board material respectively. Given $C_{FR4}\tan\delta_{FR4} <\sim 0.04$ pF, and $T_{FR4}\cong 296$ K, with a rectangular probability distribution, we estimate a value of $a_1\cong 8\times10^{-9}$ kHz$^{-1}$. When noise data are simulated with this *a priori* value for $a_1$ and reasonable values for $a_2$ included, errors in $a_0$ of about $2\times10^{-6}$ are predicted.

*Spectral Aberrations*

The transmission lines of the resistor and QVNS probes are tuned to produce a ratio spectrum that is as flat as practical, and the remaining quadratic term in the frequency response is accounted for through fitting. Despite these efforts, the fit residuals are not perfectly flat within the measurement band and small but broad spectral aberrations remain. These features are aberrations in the sense that they are not consistent with simple filtering effects and evidently have a more complex origin. Both data sets exhibit spectral aberrations of somewhat different character over the first few 100 kHz the net effect of which is a dependence of $a_0$ on the upper frequency limit of the fit as shown in Fig 6(a). The two data sets yield consistent values for $a_0$ for fitting limits above 525 kHz. Our highest fit limit of 650 kHz is set by the existence of higher-order frequency dependence which becomes dominant above that limit. For the region between 525 kHz and 650 kHz, however, there remains a noticeable fit limit dependence, which we interpret as the effect of the aberrations. The fitted value for $a_0$ changes by approximately $18 \times 10^{-6}$ over this frequency limit interval. At present, we cannot model or otherwise explicitly correct for this effect, and the simple quadratic fitting process essentially averages over the aberrant regions. We account for this ambiguity by assigning a rectangular probability distribution of half-width $18 \times 10^{-6}$ which yields a standard uncertainty of approximately $10.4 \times 10^{-6}$ to account for the existence of spectral aberrations.

*Electromagnetic Interference*

EMI is another unavoidable source of error. It is typically intermittent, caused by nearby electrical machinery not associated with the JNT, and for magnetically coupled EMI is difficult to shield. To minimize possible EMI effects, all of the analog electronics and the ADCs are operated from independent battery power supplies, and the ADCs are coupled to the computer via fiber-optic links. As described above and elsewhere [9], EMI of sufficiently large amplitude was found to overdrive the amplifiers, which produced distortion and frequency-dependent errors. Additional shielding and careful grounding reduced these overloads and measurement analysis allowed us to detect chops containing overloads so they could be removed from the final data analysis. However, EMI signals with amplitudes that are small relative to the measured signals, especially those buried within the amplifier noise, are more difficult to detect, especially in measurements of the resistor signal. While statistical tests on the averaged power spectra can detect stationary single-frequency EMI, in general the tests are not sufficiently powerful to detect all types of EMI [22].

Evidence of the absence of EMI effects in the QVNS measurements was obtained by separate measurements of the QVNS and resistor circuits, such that the two circuits generate no correlated signals. The QVNS is operated so that it generates zero volts (zero output from the code generator) and the sense resistor circuit is modified by connecting two of the differential leads to 'dummy' sensor resistors of the same 50 Ω impedance and geometric layout as the real sensor [23]. When these modified 0 V sources are measured with the JNT electronics, the resulting cross-correlated signals decrease with increasing integration period (more chops) as uncorrelated noise from their lead resistances is gradually reduced to reveal EMI signals. Any non-zero noise-power signals then indicates a potential EMI signal. After a full day of integration, no EMI signals were found in the measurement bandwidth for either source. Through these direct measurements and analysis of the EMI, we estimate an EMI uncertainty of 2 ppm.

*Distortion*

Distortion products due to any non-linearity in the correlator lead to errors in the noise power measurement, and hence in the measured spectral power ratios. The distortion in each channel can be modeled as

$$V_{i,\text{out}} = \sum_{j=0}^{\infty} a_{i,j} \left( V_{\text{in}} + V_{n,i} \right)^j \tag{3}$$

where $V_{\text{in},i}$ and $V_{\text{out},i}$ are respectively, the input and output voltages, and $V_{n,i}$ are the equivalent input noise voltage for each amplifier channel, and the linear terms $a_{1,1}$ and $a_{2,1}$ are assigned a value of 1.0. If the most significant terms only are considered, then the measured spectral power ratios, calculated from the product of the two output voltages, have the expected value

$$\left. \frac{\langle V_Q^2 \rangle}{\langle V_R^2 \rangle} \right|_{\text{meas}} = \frac{\langle V_Q^2 \rangle}{\langle V_R^2 \rangle} \left( \begin{array}{l} 1 + 3\left(\frac{a_{1,3}}{a_{2,1}} + \frac{a_{2,3}}{a_{1,1}}\right)\left(\langle V_Q^2 \rangle - \langle V_R^2 \rangle\right) + 3\frac{a_{1,3}}{a_{2,1}}\left(\langle V_{n,1,Q}^2 \rangle - \langle V_{n,1,R}^2 \rangle\right) + 3\frac{a_{2,3}}{a_{1,1}}\left(\langle V_{n,2,Q}^2 \rangle - \langle V_{n,2,R}^2 \rangle\right) \\ + \frac{a_{1,2}a_{2,2}}{a_{1,1}a_{2,1}}\left[ 3\left(\langle V_Q^2 \rangle - \langle V_R^2 \rangle\right) + \left(\langle V_{n,1,Q}^2 \rangle - \langle V_{n,1,R}^2 \rangle\right) + \left(\langle V_{n,2,Q}^2 \rangle - \langle V_{n,2,R}^2 \rangle\right) + \frac{\langle V_{n,1,Q}^2 \rangle \langle V_{n,2,Q}^2 \rangle}{\langle V_Q^2 \rangle} - \frac{\langle V_{n,1,R}^2 \rangle \langle V_{n,2,R}^2 \rangle}{\langle V_R^2 \rangle} \right] \end{array} \right) \tag{4}$$

where the equivalent input noise powers are given the additional subscript Q or R to indicate the noise source being measured. This source distinction is important because the attributed uncorrelated noise in each channel must include the thermal noise generated by the lead wires of the respective connections to the preamplifiers. Equation (4) shows that if the various noise sources are matched according to $\langle V_Q^2 \rangle = \langle V_R^2 \rangle$, $\langle V_{n,1,Q}^2 \rangle = \langle V_{n,1,R}^2 \rangle$, and $\langle V_{n,2,Q}^2 \rangle = \langle V_{n,2,R}^2 \rangle$, then in the error in the measured spectral power ratio, due to distortion effects, is zero. This is the rationale for the close matching of the QVNS and R noise sources as described in section 2.

An important feature of Eq. (3), unlike previous distortion analyses (e.g. [4, 23]), is that it includes the second-order distortion coefficients $a_{1,2}$ and $a_{2,2}$. These terms have been neglected in earlier analyses because they appear in Eq. (3) as the product of two small terms, and were therefore expected to be negligible. However, measurements using a two-tone test showed that the second-order distortion is, in fact, dominant [6, 8]. The reduction of second order distortion was also the rationale for the auto-offset-null circuits described in Section 3. All of the relative error terms included in Eq. (4) scale as voltage squared.

Measurements at 300 kHz, made at several different amplitudes, and then extrapolated to the typical input voltage for the correlator, show that both the second- and third-order terms, respectively, contribute only $2 \times 10^{-5}$ and $3 \times 10^{-6}$ error in the power spectral ratio at 300 kHz, without considering the additional benefits of matching the noise powers. However, this simple analysis neglects the potential frequency dependence of the non-linearities. Frequency dependence may arise because non-linearities can be reduced by negative feedback with the consequence that second- and third-order distortion terms of Eq. (4) respectively, rise as frequency squared and cubed as the loop gain of the amplifier falls away. In the correlator, the frequency dependence of at least some of the contributions to the second-order distortion products will therefore increase as frequency to the fourth power. When the frequency dependence is considered, such distortion effects become increasingly significant as the 650 kHz upper cutoff frequency of the correlator is approached. The greatest contribution at 650 kHz is due to the second-order distortion, but can be offset by matching the QVNS and R noise powers to better than about 0.1%, and the uncorrelated noise power in each channel to within 0.2%. The match in the QVNS and R powers are easily achieved by the appropriate choice of the factor Q in Eq 1, as already discussed. The match in the uncorrelated noise sources, or more specifically the leadwires to the preamplifiers, is achieved by the insertion of small (< 1 Ω) resistors. We assign a standard uncertainty of $1\times10^{-6}$ to account for any remaining unmatched distortion effects.

The combined uncertainty $u_r(SPR)$ for the spectral power ratio is the root-sum-square (RSS) of the individual uncertainty components shown in Table II.

Table II. The principle standard uncertainties for the measurement of the spectral power ratio.

|  | $u_r / 10^{-6}$ |
|---|---|
| Statistics | 7.4 |
| Correction Model | 2.0 |
| Spectral aberrations | 10.4 |
| EMI | 2.0 |
| Distortion | 1.0 |
| $u_r(SPR)$ | 13.1 |

## 4.2. Resistance Measurement

The resistance measurement was performed periodically between noise data acquisition cycles. A simple bipolar DC method was used to compare voltages across the two-terminal-pair junctions of the resistor-probe resistance with the analogous voltages from a calibrated 100 Ω reference resistor. Excitation currents of ±500 μA were used for most measurements and checks for power dependence indicated no observable effects at that level. The statistical relative uncertainty was typically $0.4\times10^{-6}$ for a series of 250 individual measurements.

The reference resistor is a 100 Ω hermetic encapsulated type with a calibration history traceable to the as-maintained ohm via the quantum-Hall effect (QHE) at NIST.[24] The use of the QHE to maintain the ohm is the standard approach [26] worldwide to assure the highest uniformity and stability in the disseminated unit by adopting the conventional value for the Von Klitzing constant $R_{K-90} \equiv 25,812.807$ Ω. This value is judged to be consistent with the SI ohm with an uncertainty of only $0.2\times10^{-6}$. The drift rate of the reference resistor has been determined to be 0.05 μΩ/Ω-yr. The estimated relative uncertainty in the reference resistance including its calibration and instability is $0.15\times10^{-6}$. Hence, the uncertainty in the ratio $X_R \equiv R/R_{K-90}$ is completely dominated by the uncertainty in R in the as-maintained units.

The individual metal foil resistors forming the sensing resistor $R$ are not perfectly stable under thermal cycling. These resistors will exhibit some time-dependent drift after being cooled from ambient conditions to 273.16 K in a WTP cell. The drift is manifest as an exponential relaxation starting from the original thermal cycle which presumably is due to a differential expansion strain in the foil. Measurements of the resistance are normally made only after a day or so after the R-probe is inserted into the WTP cell. We assign a relative uncertainty of $0.5 \times 10^{-6}$ to account for the effects of the resistance-relaxation effect.

The general formulation of the Nyquist formula for the mean-squared voltage fluctuations across an arbitrary impedance $Z(f)$ is governed only by the real or dissipative part $\text{Re}\{Z(f)\}$, which is $4kT\text{Re}\{Z(f)\}$ in the low frequency limit (i.e. $f \ll kT/h$).[27] The resistance is measured at DC only, so any difference between the DC value $R_{DC}$ and $\text{Re}\{Z(f)\}$ results in an error in the inferred value for $kT$.

Several physical effects need to be considered in evaluating $\Delta Z(f) \equiv \text{Re}\{Z(f)\} - \text{Re}\{Z(0)\}$. First there is the effect of ordinary but small parasitic shunt capacitance $C_R$ and series inductance $L_R$ associated with the resistor. This results in the usual modification of the form $\Delta Z(f) \cong R_{DC}(1 + 2(\tau_{LC}\omega)^2 - (\tau_{RC}\omega)^2)$ where $\tau_{RC} \equiv RC_R$ and $\tau_{LC} \equiv L_R C_R$ and $\omega = 2\pi f$. Thus, the leading terms are quadratic in frequency and so would be indistinguishable from the larger quadratic terms governed by the reactances associated with the transmission-line cabling. These types of errors are then of no consequence when the spectral ratio is already being fitted for a quadratic frequency coefficient $a_2$. Similar to the above discussion in section 4.1, any losses $\tan(\delta_{RC})$ associated with the small capacitance of the resistor foil's ceramic substrate would produce a linear frequency dependence with a coefficient $a_1 \cong -2\pi\tau_{RC}\tan(\delta_{RC})$, but we estimate this would be a factor of 10 or more smaller than that already given above. Finally, the normal departures from uniform current density, which take place as frequency increases in any conductor, will also produce frequency dependence in the measured resistance. In this case the frequency dependence is governed by the ratio $t/\delta_s$ where $t$ is a thickness of the foil and $\delta_s = (2\rho/\omega\mu)^{-1/2}$ is the so-called skin depth. It can be shown [25] that the surface impedance of a thin conducting sheet of resistivity $\rho$ is

$$Z_s = \frac{\rho}{\delta_s} \frac{(1+j)}{\tanh\left(t(1+j)/2\delta_s\right)} \cong \frac{2\rho}{t}\left[1 + \frac{2j}{3}\left(\frac{t}{2\delta_s}\right)^2 + \frac{4}{45}\left(\frac{t}{2\delta_s}\right)^4 + O\left(\frac{t}{2\delta_s}\right)^6 \ldots\right]. \qquad (5)$$

The leading real term in the power series expansion of Eqn. 5 is 4$^{th}$ order in $t/\delta_s$ which is 2$^{nd}$ order in frequency. Thus the ac resistance is again modified in the same way as in the previous case and the quadratic frequency dependence is indistinguishable from the ordinary filtering terms already being accounted for in the fit.

Another origin of finite values for $\Delta Z(f)$ are thermoelectric effects, which give rise to errors in $R_{DC}$ and rapidly diminish with increasing frequency. The effects that are first order in the Seebeck coefficient are normally canceled by current reversal. There are, however, second order thermoelectric effects which do not cancel with current reversal but rather add.[28] In order to uncover any possible sign of these 2$^{nd}$-order effects, we measured our foil resistors at DC and both 30 Hz and 90 Hz AC excitation and found that $\Delta Z(f)$ for the two test frequencies was effectively zero within the uncertainties of the measurements. We account for any remaining undetected presence of these effects with a relative uncertainty component of $0.1 \times 10^{-6}$.

Table III lists the various uncertainty components for the resistance measurement. The combined uncertainty for the resistance measurement $u(R)$ is the root-sum-square of the individual uncertainty components.

Table III. Sensing Resistance Measurement Uncertainties.

|  | $u_r / 10^{-6}$ |
|---|---:|
| Statistics | 0.40 |
| Reference Resistor | 0.15 |
| Drift | 0.50 |
| Frequency dependence | 0.10 |
| Thermoelectric effects | 0.10 |
| $u_r(R)$ | 0.67 |

### 4.3. Water Triple Point Realization

The resistance probe is directly immersed into the thermowell of a standard water triple point cell. While $T_{WTP}$ is defined in the SI as 273.16 K exactly, we can only realize the definition to within a combined uncertainty $u(T_{WTP})$ which depends on a number of limiting effects. The individual uncertainty components are listed in Table IV.

The cell is of a standard design made from borosilicate glass using distilled and de-gassed continental ground water. The pressure head correction is nominally −0.2 mK for the cell but an uncertainty of 0.05 mK is assigned to the correction due to ambiguities in the effective depth of the sensing resistor originating from the probe design.

The thermowell is filled with ethanol to promote heat transfer along the approximately 30 cm of the immersion column roughly corresponding to the height of the ice mantle. The construction of the probe utilizes a 6.3 mm diameter stainless steel sheath over most of the probe length. The interior consists of two 1.5 mm diameter tubes as grounded shields for each twisted pair of lead wires. Despite this generally conservative design, the probe's immersion characteristics are imperfect and the probe exhibits immersion errors greater than that of a standard platinum resistance thermometer.[29] A small platinum thermometer is embedded in the interior of the probe housing near the location of the sensing resistor and this has been used to evaluate the immersion error. A standard uncertainty of 0.2 mK or 0.73 µK/K is assigned to account for immersion errors.

Other uncertainties exist due to the unknown isotopic composition of the water. We assign a 0.07 mK uncertainty for this unknown variation from the defined composition [30]. The concentrations of chemical impurities is likewise unknown, but based on prior experience with cells of similar design, age and construction we assign a standard uncertainty of 0.05 mK to account for the effects of impurities.

Special provisions have been added in the most recent resistor package designs to facilitate a lower thermal impedance between the package and the inner copper block of the resistor probe. A 0.5 mm thick printed circuit board made from polyamide facilitates mounting of the sensing resistor package to a flat mounting surface on the block. This circuit board accommodates direct contact between the beryllia header and the mounting block through a 'PCB via' and mediated by a thin layer of heat transfer grease. This design helps ensure that the sensing resistor will equilibrate to the proper temperature of the cell. We assign an uncertainty of 0.05 mK associated with any remaining imperfect equilibration of the sensing resistor with the WTP cell.

Table IV. Water Triple Point Temperature realization standard uncertainties.

|  | $u_r / 10^{-6}$ |
|---|---|
| Pressure Head correction | 0.18 |
| Immersion Errors | 0.73 |
| Isotopic Variation | 0.26 |
| Chemical impurities | 0.18 |
| Thermal equilibrium | 0.18 |
| $u_r(T_{WTP})$ | 0.84 |

### 4.4. QVNS synthesis and scaling factor

The quantized nature of the voltage pulses, which are the basis for the synthesized waveforms produced by the QVNS, ensures that the uncertainties arising from the QVNS pseudo-noise signal are small. However, there is still some potential for errors and the most significant error contribution comes from undesired nonquantized current signals, which are associated with the pulse-bias signals from the digital code generator that biases the Josephson junctions in the QVNS circuit. The two primary concerns are input-output coupling, which could produce signals on the QVNS voltage leads, and voltage signals produced by bias currents driving the inductance between the junctions in the QVNS circuit. Multiple dc blocks are used on the pulse-bias leads [10] in order to reduce the unwanted signals that reside in the measurement bandwidth. Fortunately, these signals are small enough that they can be eliminated from the pulse waveform without compromising the pulse quantization. These errors are most significant at higher frequencies and have been determined to be less than 2 pV for tones below 650 kHz. Assuming these inductive signals was combined in quadrature with the Josephson voltage signals, they would contribute about $2\times10^{-9}$ to the relative uncertainty. Note that without the dc blocks, the ~4 pH geometric inductance of the Josephson array circuit would produce 130 pV signal, which in quadrature would generate a quadrature error of $\sim7\times10^{-6}$.

Another effect that can limit measurement uncertainty is variations in the voltage amplitudes of the synthesized harmonic tones for the QVNS waveform due to digitization or "quantization" noise, which is inherent in the digital-to-analog generation of desired waveform signal from the discrete high-frequency QVNS pulses. The software generating the code for the QVNS shapes this digitization noise and moves most of it to the high-frequency end of the spectrum. A third-order modulator algorithm is used to produce the waveforms for this paper, and their amplitudes at 650 kHz deviate from the target voltage by less than 5 parts in $10^7$. Calculated corrections could be applied to remove these exactly calculable variations in the waveform from the measurement [12]. Since this variation does not currently limit the measurement uncertainty, and can be removed in the future, we do not include it in the uncertainty analysis,

### 4.5. Combined Total Uncertainty

The total combined standard uncertainty in the measurement of $k$ is calculated as the root-sum-square of the principle uncertainty components. The summary is shown in Table V. The uncertainty for the Planck constant is the recommended CODATA estimate [12]. This determination, as per Eqn. 1, is for the proportionality between $k$ and $h$, hence, the only SI unit realizations required are for temperature and time. The second is realized from our local timebase $f_s$. The clock frequencies for the sampling and pulse bias source use a 5 MHz reference that is accurate to a few parts in $10^{15}$ and is traceable to the NIST primary frequency standard (NIST-F1) and others around the world.

The dominant uncertainty is from the experimental determination of the spectral power ratio and the dominant component of that uncertainty is from the spectral aberrations discussed in section 4.1.

Table V. Principle standard uncertainty components for the determination of $k$ in Eqn. 1

| component | factor | $u_r / 10^{-6}$ |
|---|---|---|
| Spectral Power Ratio | $V^2_R / V^2_Q$ | 13.1 |
| Resistance Measurement | $X_R$ | 0.67 |
| Temperature Realization | $T_{WTP}$ | 0.84 |
| Synthesis and Scaling Factor | $(QN_J)^2$ | .002 |
| Planck Constant | $h$ | 0.05 |
| Sampling Frequency | $f_s$ | <<0.001 |
| $u_r(k)$ | | 13.1 |

## 5. Conclusion

We have performed a determination of the Boltzmann constant using data derived from the comparison of Johnson noise with quantized voltage noise as synthesized via pulse-biased Josephson arrays. The Johnson noise was derived from a known resistance at a known temperature and the quantized voltage noise was derived from noise waveforms of calculable spectral density. Measurements were performed with two different reference waveforms constructed from harmonic tones of different amplitudes and tone spacings. The noise power of the harmonic tones of the reference waveforms was matched to spectral power density of the resistor. The SI kelvin and second are the only two unit realizations required for this determination.

The results are based on two measurement sets taken 3 months apart subject to different experimental conditions and different noise synthesis parameters. The only correction required was a simple fitting process to remove remnant quadratic frequency dependence in the ratio spectra. Our data yield $(k-k_{2006})/k_{2006} = +0.6(13)\times10^{-6}$ or $k = 1.380651(18)\times10^{-23}$ J·K$^{-1}$. The value determined is directly proportional to the Planck constant.

Our estimated uncertainty is dominated by systematic effects that produce aberrations in the ratio spectra and by the random statistical uncertainties achievable given the available volume of data that was accumulated over 117 h for each data set. Neither of these contributions are currently thought to represent fundamental limits on this electronic measurement of the Boltzmann constant, and further improvements in the electronics are expected to yield even lower uncertainties than those reported here.

## 6. Acknowledgements

The authors thank Charlie Burroughs for chip packaging, Norm Bergren for QVNS fabrication and technical assistance, Anthony Kos for constructing new ADC boards, Sae Woo Nam for creative ideas and years of productive collaboration, Ryan Toonen for early measurements of amplifier nonlinearities, Thijs Veening for electronics assembly, and Rand Elmquist for providing our traceability to the QHE Ohm. We also thank John Martinis for inspiring the QVNS-JNT program and for designing the original electronics, and Carl Williams for support of the NIST JNT/Boltzmann constant program. This work was performed at and support by the National Institute of Standards and Technology. It is a contribution of the U.S. government that is not subject to U.S. copyright.